\documentclass[12pt]{article}
\usepackage{amsmath,amssymb,bm,epsf,epsfig,graphicx}
\input epsf.sty
\usepackage{hyperref}
\numberwithin{equation}{section}

\newcommand{\Tr}{\mathop{\rm Tr}\nolimits}
\newcommand{\re}{\mathop{\rm Re}\nolimits}
\newcommand{\im}{\mathop{\rm Im}\nolimits}

\newcommand{\ad}{\mathop{\rm ad}\nolimits}

\def\bra#1{\langle #1 |}
\def\ket#1{|#1 \rangle}
\def\aver#1{\left\langle\, #1 \,\right\rangle}

\let\eps = \epsilon

\def \be {\begin{eqnarray}}
\def \ee {\end{eqnarray}}
\def \bdm {\begin{displaymath}}
\def \edm {\end{displaymath}}

\def\del {\partial}
\def\0{\nonumber}

\begin{document}
\vskip 2.1cm

\centerline{\large \bf A  simple solution for   marginal
deformations} \vspace{.2cm}
 \centerline{\large \bf in open string field theory}
\vspace*{.5cm}

\begin{center}

{\large Carlo Maccaferri\footnote{Email: maccafer@gmail.com}}
\vskip 1 cm
{\it Dipartimento di Fisica, Universit\'a di Torino and INFN, Sezione di Torino\\
Via Pietro Giuria 1, I-10125 Torino, Italy}
\end{center}

\vspace*{6.0ex}

\centerline{\bf Abstract}
\bigskip
We derive a new open string field theory solution for boundary marginal deformations generated by chiral currents with singular self-OPE. The solution is
algebraically identical to
the Kiermaier-Okawa-Soler solution and it is gauge equivalent to the Takahashi-Tanimoto identity-based solution. It is wedge-based and  we can analytically evaluate the
Ellwood invariant and the action, reproducing the expected results from BCFT.  By studying the
isomorphism between the states of the initial and final background a dual derivation of the Ellwood invariant is also obtained.  \vfill \eject

\baselineskip=16pt

\tableofcontents

\section{Introduction and Conclusion}
A major question in  Open String Field Theory  (OSFT) is  how the
different sets of conformal boundary conditions, in a given closed
string background, can be described by the gauge orbits of
classical solutions. Hidden in this correspondence there is the
mechanism by which OSFT is supposed to tame contact-term
singularities. In the sigma-model approach one can formally move in the space of two-dimensional boundary field theories by means of boundary interactions. However,
when interpreted as operator insertions in the world-sheet path integral of the starting background, such interactions  have notorious contact-term problems.
The advantage of  Witten's cubic open string field theory, in this regard, is that   contact-singularities can be naturally avoided by
expanding the string field in the Fock-space basis (level
truncation), thanks to the explicit
``security strips" that every Fock-space state has. However, the
level expansion  is not well fit for analytic computations. On the
other hand, with the standard wedge-based analytic methods we have
today, essentially stemming from Schnabl's original work, \cite{martin}, it is not known
how to systematically deal with contact term divergences.

Notable progress has been achieved in the case of boundary marginal deformations, \cite{bible}, in \cite{Schnabl-marg, KORZ,KO,FKP}, where
consistent ways have been devised to regularize and renormalize the contact divergences of boundary marginal operators, order by order in a
 perturbative expansion in the marginal parameter, so that an exact solution of OSFT can be defined.

More recently,
a new world-sheet mechanism for regularizing the collisions of  the marginal
operators has been put forward in \cite{id-marg} by
Inatomi, Kishimoto and Takahashi.  They analyzed an analytic
tachyon vacuum solution in the background of an
identity-based solution constructed long-ago by Takahashi and
Tanimoto (TT), \cite{TT}. They were able to analytically compute the observables  of the tachyon vacuum solution and they reproduced the
disk partition function in the marginally deformed background by computing the action, and the marginally deformed closed string tadpoles
by computing the Ellwood invariant, \cite{Ellwood}.  In their construction  the contact-term
divergences of  marginal operators are automatically
resolved by  analytically continuing the boundary marginal field along
vertical line integrals into the bulk, something which is always possible
for boundary fields coming from the chiral algebra. The
spreading in the bulk of the boundary interaction  is controlled by
a function which, in a limit, localizes to the boundary, thus
reproducing the familiar marginal deformations of \cite{bible}. This is a new, convenient way of
dealing with contact term divergences,
which doesn't require any subtraction or normal ordering.

Despite this remarkable construction, and other corollary arguments \cite{TK}, it is not possible to directly evaluate the observables of the TT solution,
because it is an identity-based string field and its action, as it stands, is not defined in a standard, known sense.

The aim of this paper is to search for a new, not identity-based,
solution which realizes the above-mentioned world-sheet regularization  of contact-term divergences and, at
the same time, has well-defined observables.  Surprisingly, by just appropriately gauge transforming the TT solution, we end up rediscovering  the
Kiermaier-Okawa-Soler (KOS) solution \cite{KOS}. For various reasons
concerning its precise world-sheet realization, \cite{NO}, the KOS
solution was believed to be able  to describe only a limited class
of marginal deformations, namely the less interesting case where
the marginal operator has regular OPE with itself and therefore
there is nothing to regulate. The world-sheet description of our
new solution  is indeed quite different from the original KOS
construction, but the identical algebraic structure allows for an
analytic --algebraic-- computation of the observables which are precisely reduced to
the tachyon vacuum observables considered and computed  in
\cite{id-marg}.  We also  take the opportunity of analyzing  the physical
fluctuations around the new solution which are explicitly constructed in terms of the degrees of freedom of the perturbative vacuum.
Starting from the similarity transformation of TT, we derive a simple world-sheet transformation which can be applied to both boundary and bulk fields.
The way bulk fields are affected by this transformation precisely accounts for the change in the closed string one-point function between the starting and the final
background. With the assumption that the $g$-function doesn't change, this gives a dual derivation of the Ellwood conjecture.

Despite the very simple algebraic structure, however, the
behaviour of the solution towards the identity is, still,
 potentially  problematic since we encounter a new, previously un-noticed, singularity which occurs when negative weight fields (such as the $c$-ghost) are placed off the boundary on a vanishing  width wedge state. We devote an
appendix to a preliminary presentation of these new kind of
singularities which would deserve, by themselves, further study and whose presence, if not properly tamed, can be quite dramatic. Luckily, it is
possible to  avoid these singularities by deforming
the original solution into a one-parameter gauge orbit which is
safe by construction and which reduces to our original solution in
a limit. Quite remarkably, the observables of the regularized solution can be exactly
shown to reduce to the difference in observables of tachyon vacuum solutions, where the regulator can be safely removed.

The  solution we are proposing is
quite handy (essentially as easy-to-handle as the original KOS solution) and at least for chiral marginal deformations is hopefully more advantageous than the standard
approaches for singular OPE's such as the  counter-terms generalizations of $B$-gauge solutions
\cite{Schnabl-marg, KORZ} or the general method of \cite{KO,
FKP}, which are  perturbative approaches in
the marginal parameter. Our construction is based on the TT solution and hence on marginal deformations,
 but the algebraic structure we describe is completely general. We thus hope our results can be a useful step towards the analytic construction
 of more general backgrounds in open string field theory, whose numerical landscape has been recently shown to be vaster than what is known analytically, \cite{KMS, ising}.

\section{From TT to KOS }

In this section we first review the needed ingredients from the Takahashi-Tanimoto (TT) solution, \cite{TT}, formulated in the sliver frame.  Then we show that, after a gauge transformation, the TT solution is mapped to a new solution which is algebraically identical to a KOS solution \cite{KOS}.

\subsection{TT Solution}

We start with a chiral current algebra
\be
\jmath^a(z)\jmath^b(0)=\frac {g^{ab}}{z^2}-\frac{c^{abc}}{z}\jmath^c(0)+(reg.),\label{curr}
\ee
and its antiholomorphic counterpart
\be
\bar\jmath_a(\bar z)\bar \jmath_b(0)=\frac {g_{ab}}{\bar z^2}-\frac{c_{abc}}{\bar z}\bar \jmath_c(0)+(reg.),\label{curr}
\ee
with totally antisymmetric structure constant $c^{abc}$.
Our reference  BCFT$_{0}$ is chosen to preserve a linear combination of the two isomorphic chiral algebras, from which it is possible  to define a single chiral current, defined on the
whole complex plane (doubling trick)
\be
 j^a(z)&=&\jmath^a(z),\quad \im z>0\\
j^a(z)&=&\Omega^{ab} \bar \jmath_b(\bar z),\quad \im z<0,
\ee
where $\Omega^{ab}$ is gluing map which is part of the data which define the starting background BCFT$_{0}$.
\be
\jmath^a(z)&=&\Omega^{ab} \bar \jmath_b(\bar z),\quad \im z=0.
\ee

The current algebra structure (\ref{curr}) guarantees that each $j^a(z)$, when placed at the boundary of
the world-sheet, generates an exactly marginal boundary
deformation of BCFT$_0$, \cite{bible}. The TT identity-based solution can then be written as a state in BCFT$_0$ as
\be
\Phi=\int_{-i \infty}^{i \infty}\frac{dz}{2\pi i}\, \left(f^a(z)
cj_a(z)+\frac12 f^af_a(z) c(z)\right).\label{nonab}
\ee
The $f^a(z)$ are functions defined on the imaginary axis, whose properties will be derived shortly.
Here we are employing the rather formal but quite useful notation \cite{Baba, id-marg} $$\phi(z)\equiv e^{z K}\phi e^{-z K},$$
 which allows to manipulate string fields as if they were local operators on the world-sheet\footnote{The well known fields $K,B,c$ are used in the conventions of  \cite{simple}.}.
 For generic $z$, $\phi(z)$ is a formal string field
 which only makes sense if it is multiplied (from the correct side) by a wedge state of minimum width $|{\rm Re} z|$. When ${\rm Re}z =0$,  $\phi(z)$ is an identity based string field
 which can be given a Fock space expansion and which can be multiplied by  wedge based states. The identity-like string field $\phi$ is defined as  $$\phi=\phi(0)\equiv \tilde\phi(1/2) I,$$ where
$\tilde\phi(w)$ is a local vertex operator in the $\frac2\pi\arctan$-sliver frame, and $I$ is the identity string field.

For concreteness we will specialize to  a single polarization inside the current algebra (\ref{curr}), by choosing one
single current
\be
j(z)&\equiv&\frac{t^a}{\sqrt{t_at_b g^{ab}}} j_a(z),\quad\quad\rightarrow\quad\quad f^a(z)=\frac{t^a}{\sqrt{t_at_b g^{ab}}} f(z),
\ee
for constant $t^a$,  with OPE
\be
j(z) j(w)=\frac1{(z-w)^2}+reg,
\ee
although most of our results readily apply to the fully non abelian case (\ref{nonab}).

With this understanding we  explicitly write
\be
\Phi=\int_{-i \infty}^{i \infty}\frac{dz}{2\pi i}\, \left(f(z) cj(z)+\frac12 f^2(z) c(z)\right).
\ee
Given a generic vertex operator
$\phi(z)$ in the sliver frame, the Fock space definition of the
identity-based string field $\Phi$ is given by computing a correlator on a cylinder $C_L$ of width
$L=1$
\be\label{Fock}
\Tr\left[\Phi e^{-\frac K2}\phi e^{-\frac K2}\right]=\int_{-i \infty}^{i \infty}\frac{dz}{2\pi i}\,\left\langle\left(f(z) cj(z+1/2)+\frac12 f^2(z) c(z+1/2)\right)\phi(0)\right\rangle_{C_1}.\label{coeff}
\ee
In order for $\Phi$ to have  well-defined Fock space coefficients (\ref{coeff},
the function $f(z)$ must vanish fast enough at the midpoint $\pm i
\infty$, so that the $dz$ integral will be finite. The finiteness
of the first term involving $cj(z)$ gives the generic condition
\be
\int_{-i \infty}^{i \infty}\frac{dz}{2\pi i} f(z)\,H(z) <\infty,
\ee
where
\be
H(z)\equiv\aver{cj(z) \phi(1/2)}_{C_1}=O(1), \quad z\to\pm i\infty,
\ee
is the contraction between $cj$ on the imaginary axis and the test state at $z=1/2$. This condition essentially states that $f(z)$ should be integrable
towards $\pm i\infty$.
The finiteness of the second term involving $c(z)$ gives a much
stronger constraint since the negative weight field $c$ must be
damped as it approaches the midpoint. For example, by
contracting with $c\del c(0)\ket0$,  we get the condition
\be
\int_{-i \infty}^{i \infty}\frac{dz}{2\pi i} f^2(z)\cos^2\pi z<\infty.
\ee
Other contractions with ghost number two Fock states similarly
imply that  $f(z)$ must separately vanish at $\pm i\infty$ at least
exponentially, faster than $e^{-\pi|z|}$, to make the integral
convergent. We will see in the appendix that the requirement of finite contractions with
generic wedge based states
will further damp the behaviour of $f$ at the midpoint.\\
Let's see how the equation of motion works in the
sliver frame. In order  to consider  $Q\Phi+\Phi^2$ as a concrete
thing, we need some world-sheet, since
 this is not provided by the solution itself. Let us then consider
 \be
 e^{-\epsilon_1 K}(Q\Phi+\Phi^2)e^{-\epsilon_2 K}.
 \ee
 The kinetic term readily gives
 \be
e^{-\epsilon_1 K}(Q\Phi)e^{-\epsilon_2 K}=\frac12\int_{-i \infty}^{i \infty}\frac{dz}{2\pi i} f^2(z) \;e^{-\epsilon_1 K} c\del c(z)e^{-\epsilon_2 K}.
 \ee
 The interaction term gives three possible contributions
\be
e^{-\epsilon_1 K}(\Phi^2)e^{-\epsilon_2 K}&=&\frac12\int_{-i \infty}^{i \infty}\frac{dw}{2\pi i}f(w)
\int_{-i \infty}^{i \infty}\frac{dz}{2\pi i}f(z)e^{-\epsilon_1 K}(cj(z)cj(w)+cj(w)cj(z))e^{-\epsilon_2 K}\0\\
&+&\frac12\int_{-i \infty}^{i \infty}\frac{dw}{2\pi i}f(w)\int_{-i \infty}^{i \infty}\frac{dz}{2\pi i}f(z)e^{-\epsilon_1 K}(cj(z)c(w)f(w)+f(w)c(w)cj(z))e^{-\epsilon_2 K}\0\\
&+&\frac18\int_{-i \infty}^{i \infty}\frac{dw}{2\pi i}f^2(w)\int_{-i \infty}^{i \infty}\frac{dz}{2\pi i}f^2(z)e^{-\epsilon_1 K}(c(z)c(w)+c(w)c(z))e^{-\epsilon_2 K}.
\ee
We now demand that $f(z)$ is analytic in an infinitesimal strip containing the imaginary axis.\footnote{This is not strictly needed, but it is a fairly general simplifying assumption.} Then, since $f$ is also suppressed at the midpoint, we can slightly shift
the $dz$ integrals on the left and on the right  of the imaginary axis, respectively for the first and second terms in the parentheses (while staying on the surface
thanks to the added strips of world-sheet).
Then the two terms in the parentheses are equivalent to a contour integral around $w$\footnote{Since we are dealing with string fields and not vertex operators,
all products must be understood to be ordered, \cite{id-marg} $$\phi_1(z)\phi_2(w)=(-1)^{|\phi_1||\phi_2|}\phi_2(w)\phi_1(z),\quad\quad \re w>\re z.$$}
\be
e^{-\epsilon_1 K}(\Phi^2)e^{-\epsilon_2 K}&=&\frac12\int_{-i \infty}^{i \infty}\frac{dw}{2\pi i}f(w)
\oint_{w}\frac{dz}{2\pi i}f(z)e^{-\epsilon_1 K} cj(z)cj(w) e^{-\epsilon_2 K}\0\\
&+&\frac12\int_{-i \infty}^{i \infty}\frac{dw}{2\pi i}f(w)\oint_w\frac{dz}{2\pi i}f(z)e^{-\epsilon_1 K} cj(z)c(w)f(w) e^{-\epsilon_2 K}\0\\
&+&\frac18\int_{-i \infty}^{i \infty}\frac{dw}{2\pi i}f^2(w)\oint_{w}\frac{dz}{2\pi i}f^2(z)e^{-\epsilon_1 K} c(z)c(w) e^{-\epsilon_2 K}.
\ee
Only the $cj$-$cj$ OPE can give a simple pole
\be
cj(z)cj(w)\sim -\frac1{z-w}\,c\del c(z),
\ee
and therefore a non vanishing result
\be
e^{-\epsilon_1 K}(\Phi^2)e^{-\epsilon_2 K}&=&-\frac12\int_{-i \infty}^{i \infty}\frac{dz}{2\pi i} f^2(z) \;e^{-\epsilon_1 K} c\del c(z)e^{-\epsilon_2 K}\0\\
&=&-e^{-\epsilon_1 K}(Q\Phi)e^{-\epsilon_2 K}.
\ee


Since the solution is identity-based, it is not possible to directly compute its observables, because they would correspond to correlators on cylinders of vanishing width. To appreciate this,
let's  compute a possible (naive) regularization of the  kinetic term by simply inserting small regulating strips, for a choice of function   $f(z)=e^{z^2}$, which is well
suppressed at the midpoint. We get
\be
\Tr[\Phi e^{-\eps_1K}Q\Phi e^{-\eps_2K}]=\frac{(\eps_1+\eps_2)^2}{64\pi^3}\left(e^{\frac{\pi^2}{(\eps_1+\eps_2)^2}}\cos\frac{2\pi \eps_1}{\eps_1+\eps_2}-1\right),\quad\quad f(z)=e^{z^2}.\label{naive}
\ee
Not only the limit $(\eps_1,\eps_2)\to 0$ does not exist, but it also wildly oscillates from $-\infty$ to $\infty$.\\
Despite the failure of a naive direct evaluation of the action, following the discussion in \cite{TT, Katsumata}, the solution is expected to describe a marginal deformation with marginal parameter given by the reparametrization invariant (see appendix A for the relation  between $f(z)$ and $F(w)$)
\be
\lambda_{\rm BCFT}\equiv \int_{-i \infty}^{i \infty}\frac{dz}{2\pi
i}\, f(z)=\int_{C_{\rm left}}\frac{dw}{2\pi i}\, F(w)\label{bcft}.
\ee
This quantity is real if the reality condition (\ref{real}) is obeyed.\\
As discussed in \cite{id-marg}, it is useful to define the matter string field\footnote{$[\cdot,\cdot]$ is the graded commutator.}
\be
J\equiv[B,\Phi]=\int_{-i \infty}^{i \infty}\frac{dz}{2\pi i}\, \left(f(z) j(z)+\frac12 f^2(z) \right),
\ee
and the deformed world-sheet hamiltonian generating horizontal translations on the cylinder $C_L$
\be
K'\equiv K+J,
\ee
whose BRST variation is given by\footnote{$\del\equiv \ad_K\equiv[K,\cdot]$.}
\be
Q(K+J)=QJ=Q[B,\Phi]=\del\Phi-[B,Q\Phi]=\del\Phi+[B,\Phi^2]=[K+J,\Phi].
\ee
The string field $K+J$ is exact in the cohomology of the shifted BRST operator
\be
K+J=(Q+{\rm {\rm {\rm ad}}}_\Phi)B\equiv Q_{\Phi\Phi}B,
\ee
where we have used the notation of \cite{EM1} for the  kinetic operator between two backgrounds $A$ and $B$
\be
Q_{AB}\phi\equiv Q\phi+ A\phi-(-1)^{|\phi|}\phi B.\label{QAB}
\ee
Generic functions of $K'$ are thus killed by $Q_{\Phi\Phi}$
\be
Q_{\Phi\Phi} F(K')=0.
\ee
The string field $F(K')$, if analytic for ${\rm Re}\, K'\geq 0$, can be geometrically understood  as a superposition of wedge-states with a path-ordered exponential integration of the chiral current, \cite{id-marg}, in much the same way as
\cite{KOS, BMT, Erler-berk}
\be
F(K')&=&\int_0^\infty dt\,{\cal F}(t)e^{-t (K+J)}\\
\Tr[F(K')e^{-\frac K2}\phi e^{-\frac K2}]&=&\int_0^\infty dt\, {\cal F}(t)\left\langle e^{-\int_1^{t+1} ds J(s)}\phi\left(\frac12\right)\right\rangle_{C_{t+1}}\label{inter}\\
J(s)&=&\int_{-i \infty}^{i \infty}\frac{dz}{2\pi i}\, \left(f(z) j(z+s)+\frac12 f^2(z) \right).\label{J(s)}
\ee
Notice however that the exponential interaction integrates the marginal current $j(z)$ over the whole bulk. This bulk (rather than  boundary) integration is what naturally regularizes
the contact term divergences between the $j$'s. The more common BCFT intuition of a renormalized boundary interaction, \cite{bible}, can be achieved
 by studying the phantom term of the solution \cite{phantom-id},
along the lines of \cite{EM1, EM2}, essentially observing that very large deformed wedges can be reparametrized to finite size while localizing the function $f(z)$ to the boundary.
Indeed, considering the scaling derivation \cite{martin} $$L^-\equiv\frac12\left({\cal L}_0-{\cal L}_0^*\right),$$ we have
\be
 L^- cj(z)&=&z\del_z cj(z)\0\\
 L^- c(z)&=&(z\del_z-1)c(z),\label{L-}
\ee
and we can easily show
\be
t^{-L^-} \Phi[f(z)]=\Phi[t f(tz)].
\ee
For $t\to\infty$ (which is the needed rescaling to bring the sliver to finite width) the support of the function $t f(tz)$ gets localized to $\im z=0$ and the bulk interaction (\ref{inter}) localizes
to the boundary, see also \cite{TK} for an almost equivalent mechanism.\\
In \cite{id-marg} it was also proven that (appropriately normalizing the space time volume)
\be
\left\langle e^{-\int_0^{t} ds J(s)}\right\rangle_{C_{t}}^{\rm matter}=\left\langle 1\right\rangle_{C_{t}}^{\rm matter}=1.\label{part}
\ee
This correlator is a regularized expression for the marginally deformed disk partition function which should therefore coincide  with the undeformed one, as it is the case.

\subsection{KOS-like solution}

Using the ingredients discussed in the previous subsection, we can
write down the solution\footnote{
This is obtained via the ``Zeze map'',  \cite{Zeze},
\be
\Phi\to\Psi\equiv
F\Phi\frac1{1+A\Phi}=(1+A\Phi)(Q+\Phi)\frac1{1+A\Phi},\label{Zeze}\ee (where
$A\equiv B \frac{1-F(K)}{K}$ and, in our case,
$F(K)=\frac1{1+K}$). Because the map is a gauge transformation it
maps solutions to solutions
$$
Q\Psi+\Psi^2=F\frac1{1+\Phi A}(Q\Phi+\Phi^2)\frac1{1+A\Phi},
$$
and it can be useful for
 turning identity-based solutions into more regular ones. It is not guaranteed, however, that the ``identity-ness'' can always be removed by gauge transformations,  the residual
 solutions of \cite{dual-L} being a counter-example.}
\be
\Psi&=&\frac1{1+K}\left(\Phi-\Phi\frac{B}{1+K'}\Phi\right)\label{KOS1}.
\ee
Although  not self-evident, this solution falls in the class of solutions studied by Kiermaier Okawa and Soler (KOS), \cite{KOS}. To see this we formally write
\be
\Phi=\sigma_L Q\sigma_R,\label{pure-gauge}
\ee
where the string fields $\sigma_{L,R}$'s obey the algebraic properties
\be
\sigma_L\sigma_R&=&\sigma_R\sigma_L=1\\
\,[B,\sigma_{L,R}]&=&0.
\ee
The expression (\ref{pure-gauge}) is precisely the pure gauge form
of the TT solution, advocated in \cite{TT}. If we assume the
existence of a logarithmic chiral field $\chi(z)$ which is a
`primitive' for $j(z)$,
\be
j(z)&=&i \del\chi(z),\\
cj(z)&=&iQ\chi(z),\\
\chi(z)\chi(w)&\sim& -\log(z-w),
\ee
 then we can  write\footnote{ As an explicit example one can take  $j=i\sqrt2\del X$
and $\chi=\sqrt2 X$, for a free boson. Notice that the
exponentials defining the $\sigma$'s are  not normal ordered (the
contact singularities are spread in the bulk).}
\be
\sigma_L&=&e^{-i\chi_f},\0\\
\sigma_R&=&e^{i\chi_f}\label{sigma},\\
\chi_f&\equiv&\int_{-i\infty}^{i\infty}\frac{dz}{2\pi i} f(z)\chi(z).\0
\ee
One can explicitly verify (\ref{pure-gauge}) by appropriately
differentiating the operator/star exponentials defining the
$\sigma$'s.
As elaborated in \cite{KT-super}, we can try to trivialize the solution $\Phi$ by making $\chi_f$ an allowed state, integrating by part
\be
i\chi_f=-\int_{-i\infty}^{i\infty}\frac{dz}{2\pi i} h(z)j(z)+i\left[\frac1{2\pi i}h(z)\chi(z)\right]_{-i\infty}^{i\infty},
\ee
where, with no loss of generality, we choose $h(z)$ as
\be
h(z)=\int_{-i\infty}^{z} d\xi\, f(\xi).
\ee
However, since $\chi(z)$ is logarithmic, the boundary term only vanishes if $$h(i\infty)=\int_{-i\infty}^{i\infty} d\xi\, f(\xi)=0.$$ The parameter defined  in (\ref{bcft}) is thus zero
if and only if the solution $\Phi$ can be trivialized. Otherwise, if $\Phi$ is non trivial,  the $\sigma$'s are  formal objects which do not belong to the state space of BCFT$_0$ (very much like
bcc operators). \\
The use of the $\sigma$'s  is nevertheless quite useful to  rewrite some of the objects we previously defined. In particular we have
\be
J\equiv[B,\Phi]=\sigma_L[B,Q\sigma_R]=\sigma_L[K,\sigma_R]=\sigma_L\del\sigma_R,
\ee
and
\be
K'&=&\sigma_LK\sigma_R\\
F(K')&=&\sigma_L F(K)\sigma_R,
\ee
which allows to rewrite (\ref{KOS1}) precisely as a KOS
solution \cite{KOS}
\be
\Psi&=&\frac1{1+K}\left(\sigma_L Q \sigma_R+Q\sigma_L\frac{B}{1+K}Q \sigma_R\right).\label{KOS2}
\ee
Notice that, differently from the original paper by KOS, the formal string fields $\sigma_{L,R}$ don't correspond to local boundary insertions of weight zero matter primaries, and their world-sheet
realization is only meaningful when a pair of them appears
\be
\langle(...)\, \sigma_L(a)\;\sigma_R(b)\,(...)\rangle_{C_L}\equiv\left\langle(...)\, e^{-\int_a^b ds J(s)}\,(...)\right\rangle_{C_L},\label{bcc}
\ee
where the non-local operator $J(s)$ is defined in (\ref{J(s)}).
 In the following, whenever possible,  we will  avoid using explicitly $\sigma_{L,R}$ and instead use the more general expression (\ref{KOS1}).
 At will, one can easily switch between the two
notations, having (\ref{pure-gauge}, \ref{bcc}) in mind. In subsection 3.3 we will elaborate more on the $\sigma$'s in presence of generic vertex operators.
 Notice also that the
auxiliary derivation $B^-\equiv\frac12\left({\cal B}_0-{\cal
B}_0^*\right)$ doesn't annihilate $\Phi$ and therefore, contrary
to the original KOS construction, the solution is not in a dressed
$B$-gauge, \cite{simple, KOS}. This matches with the expectation
that a solution for marginal deformations cannot be found in a
dressed $B$-gauge when, as is generically the case here, the
marginal field has singular OPE with itself.

As a side-comment\footnote{This possibility has been suggested by Ted Erler.}, notice that given the objects,
$\left(\sigma_{L,R}, K\right)$
one can also construct a Kiermaier-Okawa-like solution, \cite{KO}, via the substitution of the building block
$$\left[e^{\lambda V(a,b)}\right]_r\to \sigma_Le^{-(a-b)K}\sigma_R=e^{-(a-b)(K+J)},$$ where the $\lambda$ dependence in the $\sigma$'s, (\ref{sigma})(or equivalently in $J$) is realized  by choosing $f(z)=\lambda \bar f(z)$, with
$$\int_{-i\infty}^{i\infty}\frac{dz}{2\pi i} \bar f(z)=1.$$
In this case everything is already finite and directly applies to the case of a marginal field with singular self--OPE (assuming it is
local wrt all the fields in the theory, which is true if it belongs to the chiral algebra).

\section{Observables}

\subsection{Ellwood invariant}
To compute the Ellwood invariant \cite{Ellwood}, and thus the boundary state \cite{KMS},  we  use  a simple but powerful trick.
Writing the solution as, \cite{NO}
\be
\Psi=\frac{1}{1+K}\Phi\frac1{1+K'}-Q\left(\frac1{1+K}\Phi\frac B{1+K'}\right)\label{psiQ}
\ee
the Ellwood invariant is easily evaluated by inserting the $KBc$--identity
\be
[B,c]=1,
\ee
as\footnote{The notation is as follows
\be
\Tr_V[\Phi]\equiv\bra I V(i,-i)\ket\Phi,
\ee
where $\bra I$ is the bpz of the identity string field and $V$ is a weight zero bulk operator $V=c\bar c V^{\rm matter}$.}
\be
\Tr_V[\Psi]&=&\Tr_V\left[\frac{1}{1+K}\Phi\frac1{1+K'}\right]
=\Tr_V\left[\frac{1}{1+K}\Phi\frac1{1+K'}[B,c]\right]\0\\
&=&\Tr_V\left[\frac{1}{1+K}[B,\Phi]\frac1{1+K'}c\right]
=\Tr_V\left[\frac{1}{1+K}J\frac1{1+K'}c\right]\0\\
&=&\Tr_V\left[\frac{1}{1+K}c\right]-\Tr_V\left[\frac{1}{1+K'}c\right],
\ee
where, in going from the second to the third line, we have used the identity
\be
\frac{1}{1+K}J\frac1{1+K'}=\frac1{1+K'}J\frac1{1+K}=\frac1{1+K}-\frac1{1+K'}.\label{diff}
\ee
What we have obtained is precisely the difference of the invariants of the Erler-Schnabl solutions in the original background and in the background expanded around $\Phi$.
\be
\Tr_V[\Psi]&=&\Tr_V[\Psi_{TV}^{(0)}]-\Tr_V[\Psi_{TV}^{(\Phi)}]\\
\Psi_{TV}^{(0)}&=&\frac1{1+K}\left[c+Q(Bc)\right]\label{ES1}\\
\Psi_{TV}^{(\Phi)}&=&\frac1{1+K'}\left[c+Q_{\Phi\Phi}(Bc)\right].\label{ES2}
\ee
The first observable has been computed in \cite{simple}, while the second has been computed in \cite{id-marg} and shown to reproduce the closed string
tadpoles of a marginally deformed BCFT at deformation parameter $\lambda_{\rm BCFT}$ given by (\ref{bcft}). We will present an alternative derivation of this result in section 4.

Notice that all traces  in the game involve computation of correlators on cylinders of generic finite width, by the usual Schwinger parametrization of
$\frac1{1+K}=\int_0^\infty e^{-t(1+K)}$.
In addition, our algebraic derivation  is also  applicable to the regularized solution (\ref{psi-reg}) discussed in the appendix, which has the advantage of having support on wedge based
states with strictly positive width, thus avoiding the potentially problematic $t\to0$ limit in the overall Schwinger integral.

\subsection{Action}

Using a similar trick we can evaluate the action. Dropping the trivial BRST exact pieces in (\ref{psiQ}) and appropriately rotating the trace we have
\be
S[\Psi]=-\frac16\Tr[\Psi Q\Psi]=\frac16\Tr\left[\frac1{1+K}\Phi\frac1{1+K'}\Phi\frac1{1+K}\Phi\frac1{1+K'}\right].
\ee
This quantity can be in principle computed as the partition function of a wedge state with insertions and deformed/undeformed regions, with four Schwinger parameters to integrate over. This doesn't look
simple at all. But let us insert $[B,c]=1$ rightmost in the trace and, as we did for the Ellwood invariant, pull out the adjoint action of $B$ on the
other string fields in the trace
\be
S[\Psi]&=&\frac16\Tr\left[\frac1{1+K}\Phi\frac1{1+K'}\Phi\frac1{1+K}\Phi\frac1{1+K'}\right]\0\\
&=&\frac16\Tr\left[\frac1{1+K}\Phi\frac1{1+K'}\Phi\frac1{1+K}\Phi\frac1{1+K'}[B,c]\right]\0\\
&=&\frac16\Tr\left[\frac1{1+K}J\frac1{1+K'}\Phi\frac1{1+K}\Phi\frac1{1+K'}c\right]\0\\
&-&\frac16\Tr\left[\frac1{1+K}\Phi\frac1{1+K'}J\frac1{1+K}\Phi\frac1{1+K'}c\right]\0\\
&+&\frac16\Tr\left[\frac1{1+K}\Phi\frac1{1+K'}\Phi\frac1{1+K}J\frac1{1+K'}c\right].
\ee
Using (\ref{diff}) three times, we get some cancellations and we end up with
\be
S[\Psi]=\frac16\Tr\left[\frac1{1+K}\Phi\frac1{1+K'}\Phi\frac1{1+K}c\right]-\frac16\Tr\left[\frac1{1+K'}\Phi\frac1{1+K}\Phi\frac1{1+K'}c\right].
\ee
Now  recognize the BRST-exact quantities
\be
\Phi\frac1{1+K'}\Phi&=&-Q\left(\frac1{1+K'}\Phi\right)\\
\Phi\frac1{1+K}\Phi&=&Q_{\Phi\Phi}\left(\frac1{1+K}\Phi\right),
\ee
which allow to integrate by part
\be
S[\Psi]&=&-\frac16\Tr\left[\frac1{1+K}Q\left(\frac1{1+K'}\Phi\right)\frac1{1+K}c\right]-\frac16\Tr\left[\frac1{1+K'}Q_{\Phi\Phi}\left(\frac1{1+K}\Phi\right)\frac1{1+K'}c\right]\0\\
&=&-\frac16\Tr\left[\frac1{1+K}\frac1{1+K'}\Phi\frac1{1+K}Qc\right]-\frac16\Tr\left[\frac1{1+K'}\frac1{1+K}\Phi\frac1{1+K'}
Q_{\Phi\Phi}c\right],\0 \ee where we have used\footnote{In the
present case we have $Q_{\Phi\Phi}c=Qc=c\del c$, however we want
to keep as generic as possible, without assuming that
$[\Phi,c]=0$, so that we can use this derivation also for the
regularized solution described in the appendix.} \be
Q_{\Phi\Phi}F(K')&=&0.\0
 \ee
 Now we insert again $[B,c]=1$ and,
again, integrate by part the adjoint action of $B$
\be
S[\Psi]&=&-\frac16\Tr\left[\frac1{1+K}[B,c]\frac1{1+K'}\Phi\frac1{1+K}Qc\right]-\frac16\Tr\left[\frac1{1+K'}[B,c]\frac1{1+K}\Phi\frac1{1+K'} Q_{\Phi\Phi}c\right],\0\\
&=&-\frac16\Tr\left[\frac1{1+K}c\frac1{1+K'}J\frac1{1+K}Qc\right]+\frac16\Tr\left[\frac1{1+K}c\frac1{1+K'}\Phi\frac1{1+K}\del c\right]\0\\
&&-\frac16\Tr\left[\frac1{1+K'}c\frac1{1+K}J\frac1{1+K'}Q_{\Phi\Phi}c\right]+\frac16\Tr\left[\frac1{1+K'}c\frac1{1+K}\Phi\frac1{1+K'}\del' c\right]\0\\
&=&-\frac16\Tr\left[\frac1{1+K}c\frac1{1+K}Qc\right]+\frac16\Tr\left[\frac1{1+K}c\frac1{1+K'}(Qc+\Phi c)\right]\0\\
&&+\frac16\Tr\left[\frac1{1+K'}c\frac1{1+K'}Q_{\Phi\Phi}c\right]-\frac16\Tr\left[\frac1{1+K'}c\frac1{1+K}(Q_{\Phi\Phi}c- \Phi c)\right],
\ee
where, in the third line, we have defined $$\del'c\equiv \ad_{K+J}c=[B,Q_{\Phi\Phi}c],$$ and in the last two lines we have used the cyclicity of the trace, the algebraic property (\ref{diff}) as well as
\be
\frac1{1+K}\del c\frac1{1+K}&=&\left[c,\frac1{1+K}\right]\\
\frac1{1+K'}\del' c\frac1{1+K'}&=&\left[c,\frac1{1+K'}\right].
\ee
Using (\ref{QAB}) we can therefore write
\be
S[\Psi]&=&-\frac16\Tr\left[\frac1{1+K}c\frac1{1+K}Qc\right]+\frac16\Tr\left[\frac1{1+K'}c\frac1{1+K'}Q_{\Phi\Phi}c\right]\0\\
&&+\frac16\Tr\left[\frac1{1+K}c\frac1{1+K'}Q_{\Phi0}c\right]-\frac16\Tr\left[\frac1{1+K'}c\frac1{1+K}Q_{0\Phi}c\right].
\ee
The last line vanishes on account of the generic property
\be
\Tr[Q_{AB}(\phi_1)\phi_2]+(-1)^{|\phi_1|}\Tr[\phi_1 Q_{BA}(\phi_2)]=0.
\ee
Therefore the action evaluated on the solution equals
\be
S[\Psi]&=&-\frac16\Tr\left[\frac1{1+K}c\frac1{1+K}Qc\right]+\frac16\Tr\left[\frac1{1+K'}c\frac1{1+K'}Q_{\Phi\Phi}c\right]\0\\
&=&-\frac16\Tr\left[\frac1{1+K}c\frac1{1+K}c\del
c\right]+\frac16\Tr\left[\frac1{1+K'}c\frac1{1+K'}c\del c\right],\label{energy-shift}
\ee
where in the last line we have specialized to our precise case
where $[\Phi,c]=0$. Same as the Ellwood invariant, this is
precisely the difference between the action of the Erler-Schnabl
solutions (\ref{ES1},\ref{ES2}) in the original and
$\Phi$-background. Using (\ref{part}), we see that the two actions
equal each other, \cite{id-marg}, as it is expected since the
solution $\Psi$ describe a continuos family of marginal
deformations of the perturbative vacuum and must have therefore a
vanishing action. It should be noted, however, that
vanishing of the action is not an algebraic  consequence of our
derivation.
There is a reason for this: given {\it any} solution $\Phi$,
one can always construct a gauge equivalent KOS-like solution
\be
\Psi=\frac1{1+K}\left(\Phi-\Phi\frac{B}{1+K+[B,\Phi]}\Phi\right),
\ee
and  follow the computation of the energy we have just presented, to reduce it to the shift in the tachyon vacuum's energy.
Since $\Phi$ can be a generic solution, there is no reason to
expect to find a vanishing action. Therefore the algebraic form of
the KOS-type  solution we are discussing, can be useful for
generic backgrounds, not just marginal deformations.

\section{Deformed background}

We can easily describe the states and the cohomology representatives in
the new open string background described by the solution $\Psi$. In
order to do so let us first address, in our formalism, the
construction of the fluctuations  around the TT-solution
$\Phi$ itself, which was  discussed in part  in \cite{TT, KT-super}.
This will allow us to make some interesting connection with the
standard BCFT description of a marginal deformation \cite{bible} and to perform an alternative, simpler, computation of the Ellwood invariant.
Let $\Xi$ be a Fock state around the starting background $\Psi=0$
\be
\Xi=e^{-\frac K2}Ve^{-\frac K2},
\ee
where $V\equiv \tilde V(1/2) I$ is an identity-like insertion.
Since the TT solution can be written as
\be
\Phi=\sigma_LQ\sigma_R,\nonumber
\ee
this implies a star-algebra  isomorphism between the original and  the deformed states
\be
\hat \Xi&\equiv&\sigma_L\Xi\sigma_R=e^{-\frac {K'}{2}}\hat V e^{-\frac {K'}{2}}\label{iso}\\
\hat V&\equiv& \sigma_LV\sigma_R.
\ee
Notice that if the $\sigma$'s would have been allowed fields, this would just be a gauge transformation.
Explicitly, using the appropriate generalization of the Leibniz rule, \cite{EM1}, we see that the cohomology problem at the TT-background is  mapped to the cohomology problem
at the perturbative vacuum\footnote{It is important that the  formal string fields $\sigma_{L,R}$ are closed but not exact, so that the states we are discussing are not trivial. Notice the
difference wrt the left/right gauge transformations of \cite{EM1}, which are instead conventional regular string fields, typically exact but not-invertible.}
\be
Q_{\Phi\Phi}(\sigma_L\Xi\sigma_R)&=&\left(Q_{\Phi0}\sigma_L\right)\Xi\sigma_R+\sigma_L\left(Q\Xi\right)\sigma_R+(-1)^{|\Xi|}\sigma_L\Xi \left(Q_{0\Phi}\sigma_R\right)\nonumber\\
&=&\sigma_L\left(Q\Xi\right)\sigma_R.
\ee
It appears that the  dressed vertex operators $\hat V=\sigma_L V\sigma_R$ are the only objects  where a concrete definition of the $\sigma$'s is needed, (\ref{sigma})
\be
\hat V(0)=e^{-i\,\ad_{\chi_f}} V(0).\label{Vtilde}
\ee
However,  the $*$-commutator  $[\chi_f,\cdot]$ can be rewritten using only local fields (while this is not true for left or right multiplication alone). Explicitly we can write (${\rm Re} z$ ``time ordering" is understood between the string fields $\chi$ and $V$)
\be
-i[\chi_f,V]=-i\left(\int_{-i\infty+\eps}^{i\infty+\eps}-\int_{-i\infty-\eps}^{i\infty-\eps}\right)\frac{dz}{2\pi i}f(z)\chi(z)\,V(0).\label{adj}
\ee
The singular part of the OPE between $\chi$ and $V$ can consist of poles or  it can contain a logarithm (in case the OPE of $j$ with $V$ contains a single pole, as it is the case when $V$ is a $j$-primary). Other cases are excluded because $j$ belongs to the chiral algebra
and it is thus local wrt all bulk and boundary fields. When $\chi$-$V$ consists of  poles, we can close the two vertical contours
and shrink around 0
\be
-i[\chi_f,V]=-i\oint_0\frac{dz}{2\pi i} f(z) \chi(z) V(0).
\ee
Consider now a primitive for $f(z)$,
\be
f(z)=i\del g(z),
\ee
integrating by part the closed contour we get
\be
-i[\chi_f,V]=i\oint_0\frac{dz}{2\pi i} g(z) j(z) V(0), \quad\quad \chi{\textrm-} V={\textrm{pole}}
\ee
Notice that, under the assumption we are temporarily  holding ($j$-$V$ contains no simple pole) the integration constant in $g$ doesn't play any role. The constant part of $g$ enters the game only
when we transform a $j$--primary, so that $j$-$V$ is a single pole and $\chi$-$V$ is a logarithm. In this case we can assume we have already diagonalized the $j$-primaries $V$
in such a way that they are eigenstates
under the action of $j$
\be
j(z) V(0)\sim  \frac{n_V}{z}\,V(0)+(reg.),\quad\rightarrow\quad\chi(z)\,V(0)\,\sim -i n_V\,\log z \, V(0) + (reg.),
\ee
and we can write\footnote{We are also assuming that $f(i x)=f(-i x)$, which allows to easily deal with the
 unphysical cuts in the logarithm (which are an artifact of the presence of $\chi$). This condition was also implicitly used in the first of the papers \cite{TT}. Notice
that a violation of $f(ix)=f(-ix)$ would not change $\lambda_{\rm BCFT}$ as defined in (\ref{bcft}).
}
\be
&&\int_{-i\infty+\eps}^{i\infty+\eps}\frac{dz}{2\pi i}f(z)\chi(z)\,V(0)\0\\
&=&-in_V\int_{-i\infty+\eps}^{i\infty+\eps}\frac{dz}{2\pi i}f(z)\log z\,V(0)+(reg)\0\\
&\sim&-in_V\int_{+\eps}^{i\infty+\eps}\frac{dz}{2\pi i}f(z)\log |z|^2\,V(0)+ (reg),\quad (\eps\to0)
\ee
This left vertical integral (which is finite because of the integrable singularity of the log and the fall-off of $f(z)$ at $i\infty$) precisely
 cancels (together with the regular parts) against the right vertical integral in (\ref{adj}). Therefore we have
\be
-i[\chi_f,V]=0, \quad\quad \chi{\textrm-} V={\textrm{logarithm}}.
\ee
We can conveniently summarize the result as
\be
-i[\chi_f,V]&=&i\oint_0\frac{dz}{2\pi i} g(z) j(z) V(0)\\
g(z)&\equiv&-i\int_0^z\, d\xi f(\xi).\label{g}
\ee
This can be  exponentiated to give
\be
\hat V(0)=e^{-i\,\ad_{\chi_f}} V(0)=e^{i\oint_0\frac{dz}{2\pi i} g(z) j(z)}V(0).
\ee
Suppose now we want to displace $V(0)$ off the boundary
$$V(0)\to V(ix).$$
 To compute the marginal transformation, we  follow  the above derivation and, again, we have to pay attention when
$V$ is a $j$--primary.  In this case we have

\be
&&\int_{-i\infty+\eps}^{i\infty+\eps}\frac{dz}{2\pi i}f(z)\chi(z)\,V(i x)\0\\
&=&-in_V\int_{-i\infty+\eps}^{i\infty+\eps}\frac{dz}{2\pi i}f(z)\log (z-ix)\,V(ix)+(reg)\0\\
&=&-in_V\int_{-i\infty+\eps}^{i\infty+\eps}\frac{dz}{2\pi i}f(z)\left(\log \frac{z-ix}{z}+\log z\right)\,V(ix)+(reg).
\ee
When we add the contribution from the right vertical path as in (\ref{adj}), the part proportional to $\log z$ cancels exactly as before, but now there is in addition the term
\be
-i[\chi_f,V(ix)]&=&-n_V {\Big(}\int_{-i\infty+\eps}^{i\infty+\eps}-\int_{-i\infty-\eps}^{i\infty-\eps}{\Big)}\frac{dz}{2\pi i}f(z)\log \frac {z-ix}{z}\,V(ix)\0\\
&=&-n_V \oint_{{\rm cut}_{(0,ix)}}\frac{dz}{2\pi i}f(z)\log \frac{z-ix}{z}\,V(ix)\\
&=&-n_V\int_0^{ix}\frac{dz}{2\pi i}f(z) (2\pi i) V(ix)\0\\
&=&i n_V\, g(ix)\, V(ix),
\ee
where the cut has been chosen so that the overall contribution vanishes when $x\to 0$.
Therefore, also  for holomorphic bulk insertions we find
\be
\hat V(ix)&=&e^{-i\,\ad_{\chi_f}} V(ix)=e^{i\oint_{ix}\frac{dz}{2\pi i} g(z) j(z)}V(ix)\\
g(z)&\equiv& -i\int_0^{z} dz f(z)\0.
\ee

Notice that when the pole between $j$ and $V$ is at least triple, the transformation will start evaluating the derivatives of $f(z)$. This is another reason to require that $f$ is analytic around the
imaginary axis.
Assuming $f(z)$ can be holomorphically extended beyond the imaginary axis (which is typically the case), we can also write
\be
\hat V(w)\equiv e^{w (K+J)}\hat V(0)e^{-w(K+J)}=e^{w K'}(e^{-i\,\ad_{\chi_f}} V(0))e^{-wK'}=e^{i\oint_w\frac{dz}{2\pi i} g(z) j(z)}V(w).\label{trans}
\ee
As an example, we can derive how the energy momentum tensor $T(z)$ is deformed by the marginal flow induced by the solution.
We have
\be
\hat T(w)=\sigma_L T(w) \sigma_R=e^{i\oint_w \frac{dz}{2\pi i} g(z)j(z) }T(w).
\ee
Using  the $j$--$T$ OPE
\be
j(z)T(w)&=&\frac {j(w)}{(z-w)^2}+(reg.)
\ee
we get, using $i g'(w)=f(w)$
\be
\hat T(w)=e^{i\oint_w \frac{dz}{2\pi i}
g(z)j(z) }T(w)=T(w)+f(w)j(w)+\frac12f^2(w),
\ee
which agrees with \cite{KT-super}.
As a consistency check we can also compute $\hat T(w)$ by taking the deformed BRST variation of the antighost $b(w)$
\be
\hat T(w)&=&Q_{\Phi\Phi}b(w)=Q b(w)+\oint_w\frac{dz}{2\pi i}\, \left(f(z) cj(z) +\frac12 f^2(z) c(z)\right)b(w)\0\\&=&T(w)+f(w)j(w)+\frac12f^2(w).
\ee
Another simple universal example is given by
\be
\hat j(w)=j(w)+f(w),
\ee
and one can easily check that, just as $$Q(j(z))=\del(cj(z)),$$ we have
\be
Q_{\Phi\Phi}\hat j(z)=\del_z(c\hat j(z))=\del' (c\hat j(z)).
\ee
This example is also teaching us that the star algebra operator $\del' ={\rm {\rm {\rm ad}}}_{K+J}$ acts on a deformed vertex operator $\hat V(z)$ precisely as $\del_z$.

An important property of the $\hat V$'s is that, as suggested by the notation, they obey {\it the same} operator algebra as the original $V's$
\be
V_i(z)V_j(w)&=&c_{ijk}(z-w)\,V_k(w),\\
\hat V_i(z)\hat V_j(w)&=&c_{ijk}(z-w)\,\hat V_k(w),
\ee
as can be directly verified from (\ref{trans}) by picking up residues in explicit examples. This also implies that traces involving deformed wedges and the $\hat V$'s will be
(up to a possible universal constant) the same as the corresponding traces of undeformed wedges and the $V$'s
\be
\Tr[e^{-t_1 K'}\hat V_1...e^{-t_nK'}\hat V_{t_n}]=\frac{g'}{g}\Tr[e^{-t_1 K}V_1...e^{-t_nK}V_{t_n}]\label{same}.
\ee
The constant $\frac{g'}{g}$ is the ratio of the traces of the deformed  and undeformed wedges, which, as proven in \cite{id-marg}, is equal to 1.

It is interesting to extend the marginal transformation (\ref{trans}) to closed-string bulk operators. In our doubling-trick notation a bulk operator will be written as
\be
V_{ij}(w,\bar w)=V_i(w)V_j( w^*),   \quad\quad  w^*\equiv \bar w,\;\im w>0,
\ee
where both $V_i$ and $V_j$ are holomorphic (but typically not chiral) fields. We thus have
\be
\hat V_{ij}(w,\bar w)=\left(e^{i\oint_w\frac{dz}{2\pi i} g(z) j(z)}V_i(w)\right)\left(e^{i\oint_{ w^*}\frac{dz}{2\pi i} g(z) j(z)}V_j( w^*)\right),
\ee
with $g(z)$ defined in (\ref{g}).
Let us now assume that both $V_i$ and $V_j$ are $j$--primaries (all boundary states obtained from BCFT$_0$ by deforming with $j$, will be written as a sum of Ishibashi states
of $j$-primaries, defined with the appropriate deformation of the gluing map, \cite{bible}).  With no loss of generality we can write down the OPE,  \cite{bible}\footnote{As an example, in case of $j=i\sqrt2 \del X$, with Neumann boundary conditions,
we have that bulk momentum modes have $a_i=-b_j$ while bulk winding modes have $a_i=b_j$. The situation is exactly opposite in case of Dirichlet boundary conditions. }
\be
j(z) V_{ij}(w,\bar w)\sim \left(\frac{a_i}{z-w}-\frac{b_j}{z-\bar w}\right)V_{ij}(w,\bar w),
\ee
from which we easily get\footnote{With our assumption $f(z)=f(-z)$ we have that $g(ix)=-g(-ix)$. Thus, with this condition, a bulk operator with $a_i=-b_j$ is not transformed by the marginal
deformation. But in fact a bulk operator with $a_i=-b_j$ has a vanishing tadpole in BCFT$_{0}$, and this remains true by deforming with $j$, \cite{bible}.}
\be
\hat V_{ij}(i x,-i x)&=&e^{i\left(a_i g(ix)-b_jg(-ix)\right)}V_{ij}(i x,-i x)\0\\
&=&e^{i(a_i+b_j) g(ix)}V_{ij}(i x,-i x).\label{ishi-trans}
\ee
Now imagine we want to compute the Ellwood invariant of the tachyon vacuum solution (\ref{ES2}), as it was done in \cite{id-marg}. After standard string field manipulations, we end up with the following
trace
\be
\Tr_{{\cal V}_{ij}}[e^{-(K+J)}c]=\lim_{x\to\infty}\left\langle  e^{-\int_{0}^1ds\,J(s)}\, c(0)c\bar c V_{ij}(i x,-i x)\,\right\rangle_{C_1},
\ee
where ${\cal V}\equiv c\bar c V$.
We can follow (and generalize to finite $x$, still assuming $f(iy)=f(-iy)$) the explicit computation of \cite{id-marg} to find
\be
\Tr_{{\cal V}_{ij}}[e^{-(K+J)}c]&=&\lim_{x\to\infty}e^{-i\pi(a_i+b_i)\lambda(x)}\Tr_{{\cal V}_{ij}}[e^{-K}c]\label{over}\\
\lambda(x)&\equiv&\int_{-ix}^{ix}\frac{dz}{2\pi i} f(z).
\ee
Or we can proceed differently, (\ref{same})
\be
\Tr_{{\cal V}_{ij}}[e^{-K'}c]&=&\lim_{x\to\infty}\Tr[\sigma_L e^{-K}\sigma_Rc{\cal V}_{ij}(ix,-ix)]\0\\
&=&\frac{g'}{g}\lim_{x\to\infty}\Tr[e^{-K}c\, \sigma_R{\cal V}_{ij}(ix,-ix)\sigma_L]=\Tr_{\check{\cal V}_{ij}}[e^{-K}c].
\ee
The closed string state $\check{\cal V}_{ij}$ is the inverse of the transformation (\ref{ishi-trans})
\be
\check{\cal V}_{ij}(ix,-ix)=\sigma_R{\cal V}_{ij}(ix,-ix)\sigma_L=e^{i\oint_{\pm ix}\frac{dz}{2\pi i}g(z)j(z)}{\cal V}_{ij}(ix,-ix)=e^{-i\left(a_i+b_j\right) g(ix)}{\cal V}_{ij}(i x,-i x).\0
\ee
Therefore we get
\be
\Tr_{{\cal V}_{ij}}[e^{-K'}c]=e^{-i\left(a_i+b_j\right) g(ix)}\Tr_{{\cal V}_{ij}}[e^{-K}c],
\ee
which coincides with (\ref{over}), remembering that we are taking $f(ix)=f(-ix)$ and  $$g(ix)=-i\int_0^{ix}dz\,f(z)=\pi \lambda(x).$$
Notice that, in this `dual' derivation, the Ellwood invariant is precisely reduced to a deformed closed string tadpole, in the sense of \cite{bible} (see e.g. eq (3.3) there), and Ellwood conjecture is
 transparent. In the BCFT description of \cite{bible},
the countours,  encircling the bulk operator, were originally at the boundary, while in this peculiar OSFT description they originate from vertical line integrals, (\ref{adj}).

Assuming that the contour integral  $\oint \frac{dz}{2\pi i} g(z)j(z)$ is well defined on local vertex operators (which is true if $j$ belongs to the chiral algebra, but generically false if $j$ is only  self-local, \cite{bible}), the isomorphism (\ref{iso}) can be performed on the whole
Fock space of BCFT$_0$ and,
 being a similarity transformation,
it is clearly compatible with both the
star product and the BRST differential.\\
All the above can be straightforwardly extended to the KOS-like solution $\Psi$ (\ref{KOS1}, \ref{KOS2}), where the previous isomorphism is dressed with the gauge parameters connecting the TT solution with the KOS solution
\be
e^{-\frac {K}{2}} V e^{-\frac {K}{2}}&\to&(1+A\Phi)\,e^{-\frac {K'}{2}}\hat V e^{-\frac {K'}{2}}(1+A\Phi)^{-1}\0\\
&=&\left(1+\frac{B}{1+K}\Phi\right)e^{-\frac {K'}{2}}\hat V e^{-\frac {K'}{2}}\left(1-\frac{B}{1+K'}\Phi\right)\0\\
&=&\left(\sigma_L-\frac{B}{1+K}Q\sigma_L\right)e^{-\frac {K}{2}} V e^{-\frac {K}{2}}\left(\sigma_R-\frac{B}{1+K}Q\sigma_R\right).\label{KOS-states}
\ee
Again, we can use generic states of  BCFT$_{0}$, to describe the off-shell degrees of freedom around the new background $\Psi$, and the cohomology is again mapped in the cohomology.\\
Notice that, with this construction of the off-shell fluctuations, the action in the new
background $\Psi$  $coincides$ with the action around the TT background $\Phi$. Notice however that the states (\ref{KOS-states}) are not real despite the almost
 reality of the solution $\Psi$ 
  $$\Psi\to \sqrt{1+K}\Psi\frac1{\sqrt{1+K}}\equiv \Psi^{\rm real}.$$ Given $\Psi^{\rm real}$ one can find real cohomology elements by using the right gauge transformation \cite{EM1}
$$U=\sqrt{F}\frac1{1+\Phi A},\quad\quad F(K)=\frac1{1+K},\;A=B\frac{1-F}{K}$$ and the (reality-conjugate)  left gauge transformation $$U^\dagger=\frac1{1+A\Phi}\sqrt F,$$ both connecting $\Phi$ with $\Psi^{\rm real}$. This gives a construction of the cohomology
which is essentially the one considered in \cite{Schnabl-marg, Ell-sing}
\be
e^{-\frac {K}{2}}c V e^{-\frac {K}{2}}\to U\,e^{-\frac {K'}{2}}c \hat V e^{-\frac {K'}{2}}\,U^{\dagger}.
\ee
This is a map from cohomology to cohomology but, contrary to the non-real construction (\ref{KOS-states}), it is not a star algebra homomorphism. It is also
possible, at least formally, to connect $\Phi$ with
$\Psi^{\rm real}$ with a {\it real} gauge transformation $W=U^\dagger\frac1{\sqrt{UU^\dagger}}$ obeying $W^\dagger W=W W^\dagger=1$, \cite{Ted-marg}
 which gives a
star algebra homomorphism compatible with reality. However, we do not see obvious problems in using the simpler non real deformed states (\ref{KOS-states}).

\section*{Acknowledgments}
 I thank  Ted Erler for  collaboration on previous unpublished results and for useful discussions and comments on a draft. I thank Isao Kishimoto and Tomohiko Takahashi
 for discussions. I thank Martin Schnabl for detailed comments on the first arxiv version.
 I thank the organizers of the String Theory workshop in Benasque 2013 for providing an inspiring environment where this project started. The research of the author is fully supported by a {\it Rita Levi Montalcini} grant.

\appendix
\section{TT on the upper half plane, BPZ and reality}
To make contact with the original form of the TT solution, we relate the function $f(z)$  to the
function $F(w)$ appearing in the work of TT
\cite{TT} by mapping the semi-infinite
cylinder $C_L$  of circumference $L=2$ (with coordinate $z$) to the upper half plane  (with coordinate $w$) by $w=\tan\frac{\pi z}{2}$.
\be
\Phi&=&\int_{-i \infty}^{i \infty}\frac{dz}{2\pi i}\, \left(f(z) c j(z)+\frac12 f^2(z) c\left(z\right)\right) \0\\
&=&\int_{-i \infty}^{i \infty}\frac{dz}{2\pi i}\, \left(f(z) \tilde c\tilde j\left(z+1/2\right)+\frac12 f^2(z) \tilde c\left(z+1/2\right)\right) I\0\\
&=&\int_{C_{\rm left}}\frac{dw}{2\pi i}\left(F(w)cj(w)+\frac12 F^2(w)c(w)\right) I\\
f(z)&=&\frac\pi2\frac{F\left(\tan\frac\pi2\left(z+\frac12\right)\right)}{\cos^2
\frac\pi2\left(z+\frac12\right)},\\
 F(w)&=&\frac2\pi\frac{f\left(\frac2\pi \tan^{-1}\frac{w-1}{w+1}\right)}{w^2+1},
\ee
where $C_{\rm left}$ is the semicircle in the complex plane connecting $-i,1,i$, oriented towards $i$.

In \cite{TT} and in the papers that followed, the authors also require that $F(w)$  obey
$$F\left(-\frac1w\right)=w^2F(w),$$ which, in the sliver frame, translates into
\be
f(z)=f(z-1).\0
\ee
We do not require this periodicity condition because (for example) a
simple scale transformation in the sliver frame (a
reparametrization generated by $L^-\equiv\frac12\left({\cal
L}_0-{\cal L}_0^*\right)$) would  not respect it. As
explained in \cite{TT, KT-super}, this property ensures that
\be\Phi_L I&\equiv&\int_{C_{\rm left}}\frac{dw}{2\pi i} \left(F(w)cj(w)+\frac12 F^2(w)c(w)\right)\,I\0\\&=&
\int_{C_{\rm right}}\frac{dw}{2\pi i} \left(F(w)cj(w)+\frac12
F^2(w)c(w)\right)\,I\equiv \Phi_R I,\label{ted}\ee
so that we can write
\be
\Phi_L I*\Phi_L I= (-1)^{|\Phi|} \Phi_R\Phi_L  I* I=\Phi_L \Phi_R  I=\Phi_L^2 I\label{oper-eom},
\ee
where the commutation between left and right charges holds if $F(\pm i)=0$.
However, to prove the
equation of motion, as we  saw in section 2, we only used that $F(w)$ vanishes at
the  midpoint and that it is analytic in an infinitesimal neighborhood of  $C_{\rm left}$.  The
corresponding right charge $\Phi_R$ can be defined, if needed,  by the same expression  (\ref{ted}) but with
$$F(w)\to \frac1{w^2}F(-1/w),$${\it i.e.} $$\Phi_R\to ({\rm bpz} \Phi_L)$$ which is a right-type charge which also vanishes at the midpoint and
is analytic around $C_{\rm right}$. In this way (\ref{oper-eom}) is still satisfied
\be
\Phi_L I*\Phi_L I= (-1)^{|\Phi|} ({\rm bpz}\Phi_L)\Phi_L  I* I=\Phi_L ({\rm bpz}\Phi_L)  I=\Phi_L^2 I,
\ee
because we can use the generic properties
\be
\Phi_L I&=&({\rm bpz}\Phi_L) I\\
A*(\Phi_L B)&=&(-1)^{|A|\,|\Phi|}(({\rm bpz} \Phi_L)A)*B,
\ee
which encode the gluing conditions
$$-1=w^{(i)}w^{(i+1)}{\Big |}_{|w^{(i)}|=1, (-1)^i{\rm Re}w^i>0},$$
for $N$--strings vertices.
If we like, given $F(w)$ defined on $C_{\rm left}$ ($\re w>0$) we can always extend $F(w)$ on $C_{\rm right}$ ($\re w<0$) by $$F(w)\equiv1/w^2 F(-1/w),\quad {\rm for}\;  {\rm Re}w<0,$$
but this isn't in general an analytic continuation.\footnote{
I thank Ted Erler for a useful discussion on this.} Since, to define the solution, we only need to know $F(w)$ on $C_L$, we avoid talking about the value of $F(w)$ on $C_R$.

The reality condition, on the other hand, gives a real constraint on $F(w)$. The string field $\Phi$ is real (bpz$=$hc) if the function  $F(w)$ satisfies \footnote{The reality condition for a related identity based solution for the tachyon vacuum has been discussed in \cite{Zeze-real}.}
\be
F(w)=\frac1{w^2} F^*\left(\frac 1w\right),\quad |w|=1, \re w>0\quad\quad \verb"Reality",
\ee
which in the sliver frame translates into the quite intuitive
\be
f(z)=f^*(-z)=f^*(z^*),\quad \re z=0.\label{real}
\ee

\section{A new singularity towards the identity}

The simple algebraic derivation of observables we have presented in section 3
is  potentially endangered by a  singularity towards the identity
which has to do with the $c$- ghost, as we now briefly explain.

To start with, it is better to specialize a bit on the function $f(z)$ which defines the solution. Because of the omnipresence of the quantity $\frac1{1+K}$,
 a basic requirement is that, when we add $\Phi$ to $K,B,c$,
 its contraction  is well defined against  wedge-based states of arbitrarily small width.
While this was essentially guaranteed in previous enlargements of
the $KBc$ algebra, \cite{KOS, BMT}, which dealt with boundary insertions, and even \cite{id-marg}, where only matter
operators were allowed to enter the bulk, here the story is more delicate. To appreciate the problem consider the simple overlap
\be
\Tr[\Phi\Omega^t c\del c]&=&\frac12\int_{-i\infty}^{i\infty}\frac{dz}{2\pi i} f^2(z) \langle c(z+t) c\del c(0)\rangle_{C_t}\\
&=&-\frac{t^2}{2\pi^2}\int_{-i\infty}^{i\infty}\frac{dz}{2\pi i} f^2(z) \sin^2\left(\frac{\pi z}{t}\right).
\ee
This integral is divergent for small enough $t>0$ unless the function $f(z)$ is suppressed at $i\infty$ more than exponentially.
For example, the standard choice by Takahashi and Tanimoto \cite{TT},
\be
F_{\rm TT}(w)=1+\frac1{w^2}\quad\quad\rightarrow\quad\quad f_{\rm TT}(z)=\frac{2\pi}{\cos^2 \pi z}
\ee
does not respect this  property. Indeed, although the TT-solution based on $f_{\rm TT}$ is finite in the Fock space, a finite $L^-$ reparametrization of it, (\ref{L-}),
 appears to be
singular\footnote{This singularity is absent if the marginal field $j$ has finite OPE with itself, which reflects in the absence of the $c$--part of the solution $ \frac12\int dz f^2(z) c(z)$}.
In particular
\be
\langle{\rm Fock}|t^{L^-}\Phi_{f_{\rm TT}}\rangle=\infty, \quad\quad t\leq \frac12.
\ee
This is certainly un-welcome for the purpose of enlarging the  $K,B,c$ algebra with $\Phi$, as we have been doing in the
previous sections. Therefore we would like to limit the choice of $f$ in such a way
that generic contractions with wedge based states and finite $L^-$ reparametrizations   give finite results.
 A simple example that does the job is the family of gaussians
\be
f_t(z)\equiv 2\lambda\sqrt{\pi}\,t\,e^{(tz)^2},
\ee
for which we have
\be
\lambda=\int_{-i\infty}^{i\infty}\frac{dz}{2\pi i} f_t(z).
\ee
In the following we will specialize to the family of gauge equivalent solutions described by $f_t(z)$.
These solutions are all related by $L^{-}$--reparametrizations
\be
\Phi_{f_t}=t^{-L^-}\Phi_{f_{t=1}}.
\ee
We can easily check that, for this choice of function, the TT solution $\Phi$ is
finite in the Fock space, and also against generic wedge states with insertions whose width can be taken
arbitrarily small. In particular, for example
\be
\Tr[\Phi_{f_{t=1}}\Omega^s c\del c]&=&-\frac{2\lambda^2s^2}{\pi}\int_{-i\infty}^{i\infty}\frac{dz}{2\pi i} e^{2z^2} \sin^2\left(\frac{\pi z}{s}\right)\0\\
&=&\frac{\lambda^2 s^2}{2\sqrt2\pi^{3/2}}\left(e^{\frac{\pi^2}{2 s^2}}-1\right).
\ee
Notice however that, although the overlap is finite for $s>0$, it nevertheless diverges super-exponentially
in the identity limit $s\to 0$. Sticking to this example, this means that the following
overlap is badly divergent
\be
\Tr\left[\Phi\frac1{1+K}c\del c\right]=\int_{0}^{\infty}dt e^{-t}\Tr\left[\Phi\Omega^t c\del c\right]=\infty.
\ee
 We may hope that further suppressing $f$ at the midpoint could improve the situation, but in fact there is a more basic problem.
When two $c$-ghosts have a separation with a tiny imaginary part on a cylinder of width $t$, the correlator always diverges in the limit $t\to 0$.
\be
\lim_{t\to0}\Tr[c(i x)e^{-tK}c\del c]=\lim_{t\to0}\frac{t^2}{\pi^2}\sinh^2\frac{\pi x}{t}=\infty,\quad \re x\neq 0
\ee
Notice that the negative scaling dimension of $c$ would suppress the correlator, but this comes together with a
rather violent exponential divergence, which only occurs
when $c$ is placed off the boundary.
Therefore, even
in the original $K,B,c$ algebra we have the problem
\be
\Tr[c(ix)\frac1{1+K}c\del c]=\int_{0}^{\infty}dt e^{-t}\frac{t^2}{\pi^2}\sinh^2\frac{\pi x}{t}=\infty,\quad \re x\neq 0.
\ee
This is a new kind of identity-like singularity which would be worth studying by itself. Notice in particular that naive attempts to evaluate the action of the TT solution $\Phi$
are affected by this singularity, (\ref{naive}). Notice that the singularity is much more violent than previous
identity-like singularities
studied in the sliver frame, \cite{dual-L},  whose behavior is typically power law. Since we  know quite little about these singularities and how they effectively cancel in the
algebraic computations we have been doing in the main text,
our primary aim will be to show that these singularities can be avoided by an infinitesimal deformation of our solution.

\section{Regularization}

The singular expressions we met in the previous section  are  structurally quite close to the expressions that appear in the computation of observables
in the main text.
For example, consider the kinetic term of the string field $\chi\equiv\frac1{1+K}\Phi$, which is a part of our solution $\Psi$, (\ref{KOS1}).
The explicit computation
goes as follows
\be
\Tr\left[\chi Q\chi\right]&=&\frac14\int_{-\infty}^{\infty}\frac{dx\,dy}{(2\pi )^2}f^2(ix)f^2(iy)\Tr\left[\frac1{1+K}c(ix)\frac1{1+K}c\del c(iy)\right].
\ee
If we consider the kernel
\be
\Tr\left[\frac1{1+K}c(ix)\frac1{1+K}c\del c(iy)\right]=-\frac1{\pi^2}\int_0^\infty dt\, e^{-t}t\int_0^1dq\sin^2\pi\left(q+i\frac{x-y}{t}\right),
\ee
we see that this is not a well defined quantity since, if we perform the $dt$ integral first, we encounter a bad exponential singularity at $t\to0$ of the
type $e^{2\pi\frac{|x-y|}{t}}$. On the other hand, performing the $q$ integral first, the dependence on $(x-y)$ drops and everything is finite.
In this case, it is not difficult to realize that the `correct' prescription for computing the above integral would be to define
\be
\Tr\left[\frac1{1+K}c(ix)\frac1{1+K}c\del c(iy)\right]=\lim_{\eps\to 0^+}\int_\eps^\infty dt e^{-t}\int_0^t ds\Tr[\Omega^sc(ix)\Omega^{t-s}c\del c(iy)].
\ee
With this regularization of the trace (cut-off in the overall Schwinger parameter of the string field whose trace we want to compute), the algebraic derivations of the Ellwood invariant
and the kinetic term, presented in section 3, are rigorously justified, as it is easy to check.
 However it is very difficult to understand this regularization  at the level of the individual string fields before $*$-multiplication and, importantly, to understand how the equation of motion in
 the action
 is violated and how (and if) it is restored when the regulator is removed.

Our algebraic computation suggests there exists a prescription (which is consistent with the equation of motion)
to correctly compute the observables. But to make this precise, we need a  regularization which allows us to control the  $t\to 0$ limit in the overall
Schwinger integral and to maintain, at the same time, the equation of motion.
Perhaps the simplest and safest approach (but other strategies might be possible) is to
realize  that the  solution we are dealing with can be obtained as
a limit of a one parameter family of gauge equivalent solutions
which have, generically, a minimum fixed width and therefore, by
construction, cannot have any singularity related to the identity
string field.
 To construct such a family is  easy and amounts to
choosing a security strip in the Zeze map (\ref{Zeze})  given by (for example, other choices are of course possible )
\be
F_\eps=\frac{\Omega^\eps}{1+\bar\eps K},\quad\quad \bar\eps\equiv 1-\eps,
\ee
where
\be
\Omega\equiv e^{-K}=\ket0_{SL(2,R)}.
\ee
The regularized solutions are given by
\be
\Psi_\eps=\frac{\Omega^\eps}{1+\bar\eps K}\Phi\left(1-\frac{B}{1+h_\eps J}h_\eps\Phi\right),\label{psi-reg}
\ee
where
\be
h_\eps=\frac{1-\frac{\Omega^\eps}{1+\bar\eps K}}{K}=\frac1{1+\bar\eps K}\left(\bar\eps+\int_0^\eps dt \Omega^t\right).
\ee
These solutions span a gauge orbit interpolating from KOS
($\eps=0$) and the generalization of the Schnabl-KORZ solution,
\cite{Schnabl-marg, KORZ}, ($\eps=1$) which, for completeness,
reads
\be
\Psi_1&=&\Omega\Phi\left(1-\frac{1}{1+\frac{1-\Omega}{K}J}\frac{1-\Omega}K B\Phi\right)\nonumber\\
&=&\Omega\Phi\left[1-\sum_{n=0}^{\infty}\left(\int_0^1 dt\Omega^t J\right)^n\int_0^1 dt\Omega^tB\Phi\right].
\ee
The strategy is therefore to define the observables of the KOS solution  as the $\eps\to0^+$ limit of the observables of the interpolating solutions (which, by gauge invariance, will be $\epsilon$-independent).
At finite $\epsilon$ it is guaranteed that no identity-singularity can affect the computation. However the price we pay for manifest regularity is that the generic solution in the orbit is
much more complicated than the original KOS solution and it is not clear, at this stage, how one could  compute the observables as we did in the main text.
But in fact we can {\it rewrite} the regularized solution (\ref{psi-reg}) again as a KOS solution, where the fields $(K,B,c,\Phi)$ have undergone the automorphism \cite{simple-ted}
\be
c&\to& c_\eps=c \frac {KB}{G_\eps(K)}c\label{aut-1}\\
B&\to& B_\eps=B \frac{G_\eps(K)}{K}\\
K&\to& K_\eps=QB_\eps=G_\eps(K)\\
\Phi&\to&\Phi\\
J&\to& J_\eps=[B_\eps,\Phi].\label{aut-2}\\
\ee
where $G_\eps(K)$ is defined by\footnote{$G_\eps(K)=K_\eps$ is a purely formal string field which is proportional to the inverse wedge $e^{\eps K}$. However it always appear in the combination
$\frac1{1+K_\eps}$ which is fine. Similar considerations apply to $B_\eps$ and $J_\eps\equiv[B_\eps,\Phi]$ which are formal by themselves but always appear in the regular combinations
$\frac {B_\eps}{1+K_\eps}$, $\frac1{1+K_\eps+J_\eps}$,  $\frac1{1+K_\eps+J_\eps}J_\eps\frac1{1+K_\eps}$, $B_\eps c_\eps$, $c_\eps B_\eps$, etc...}
\be
\frac1{1+K_\eps}=\frac1{1+G_\eps(K)}=\frac{\Omega^\eps}{1+\bar\eps K}.
\ee
Notice that the TT solution remains invariant under the automorphism.

With the new variables, and some standard algebra, we can  re-write (\ref{psi-reg}) in few interesting ways
\be
\Psi_\eps&=&\frac{1}{1+K_\eps}\Phi\frac1{1+K_\eps+J_\eps}-Q\left(\frac1{1+K_\eps}\Phi\frac {B_\eps}{1+K_\eps+J_\eps}\right)\\
&=&\frac{1}{1+K_\eps}\Phi\frac1{1+K_\eps+J_\eps}+\frac1{1+K_\eps}\Phi\frac1{1+K_\eps+J_\eps} (K_\eps+\Phi B_\eps)\0\\
&=&\frac{\Omega^\eps}{1+\bar\eps K}\Phi\left(\frac1{1+h_\eps B\Phi}\,\frac{\Omega^\eps}{1+\bar\eps K}\,\frac1{1+\Phi B h_\eps}\right)\0\\
&&+\frac{\Omega^\eps}{1+\bar\eps K}\Phi\frac1{1+ h_\eps B \Phi}\left(1-\frac{\Omega^\eps}{1+\bar\eps K}\,\frac1{1+\Phi B h_\eps}\right)\\
&=&\!\frac{\Omega^\eps}{1+\bar\eps K}\Phi\!\left(\frac1{1+h_\eps B\Phi}\,\frac{\Omega^\eps}{1+\bar\eps K}\,\frac1{1+\Phi B h_\eps}\right)\0\\&&-
Q\left(\frac{\Omega^\eps}{1+\bar\eps K}\,\frac{\Phi h_\eps B}{1+\Phi h_\eps B}\right).\label{reg}
\ee
The reader can explicitly verify that the first `physical' term in the regularized solution (\ref{reg}) has  support on wedge-based states of minimum width $2\eps$  while the
 BRST exact one has minimum width $\eps$.
Notice that we have
\be
\frac1{1+K_\eps+J_\eps}=\frac1{1+ h_\eps B \Phi}\frac{\Omega^\eps}{1+\bar\eps K}\,\frac1{1+\Phi B h_\eps},
\ee
which reveals that the automorphism mixes the objects in the game
in a rather non trivial way. In particular
\be
Q_{\Phi\Phi}\frac1{1+K_\eps+J_\eps}&=&0\\
{\ad_{K+J}}\frac1{1+K_\eps+J_\eps}&=&[Q_{\Phi\Phi},\ad_B]\frac1{1+K_\eps+J_\eps}=0,
\ee
which trivially descend from the automorphism, but which appear rather surprising in the original variables.
Other notable quantities are given by
\be
\frac{B_\eps}{1+K_\eps}&=&B h_\eps\\
\frac{B_\eps}{1+K'_\eps}&=&Bh_\eps\frac1{1+Jh_\eps}=B\frac1{1+h_\eps J}h_\eps\\
c_\eps B_\eps&=&cB\\
B_\eps c_\eps&=&Bc\\
c_\eps K_\eps B_\eps c_\eps&=&cKBc\\
c_\eps (K_\eps +J_\eps)B_\eps c_\eps&=&c(K+J)Bc,
\ee
notice that the automorphism doesn't increase the minimum width of the above quantities, which all continue to have a non vanishing support on the identity.
 Since the fields $(K_\eps,B_\eps,c_\eps,\Phi)$ have identical
properties to $(K,B,c,\Phi)$ the computations for $\Psi_\eps$ can
be read-off from the main text by formally substituting
$(K,B,c,J)$ with $(K_\eps,B_\eps,c_\eps,J_\eps)$. By inspecting the involved correlators, we see than only well defined combinations of the
deformed variables explicitly appear, so we have just to  trace back
how the simplifications in the deformed variables occur in the original variables. In doing this we encounter very non trivial simplifications
between different structures,
which would have been practically impossible to discover if not guided by the formal automorphism (\ref{aut-1}---\ref{aut-2}).
 For fixed $\eps\neq0$ we can then precisely
show  that the observables of the regularized solution $\Psi_\eps$ reduce to the shift in the observables of the tachyon vacuum's
\be
\Psi_{TV,\eps}^{(0)}&=&\frac1{1+K_\eps}(c_\eps+Q(B_\eps c_\eps))\\
\Psi_{TV,\eps}^{(\Phi)}&=&\frac1{1+K'_\eps}(c_\eps+Q_{\Phi\Phi}(B_\eps c_\eps)).
\ee
When we take the $\eps\to0$ limit these two solutions become the Erler-Schnabl solutions (\ref{ES1},\ref{ES2}), which are manifestly safe from the
identity singularities we encountered
in the previous section, simply because no explicit $\Phi$ enters in their definition and therefore there is no $c$--field going off the boundary.
For completeness, it would be interesting to have an analytic computation of the observables of $\Psi_{TV,\eps}^{(\Phi)}$, at finite $\eps$, which we leave for the future.

\end{document}